# Electronic Correlations Control Interlayer Coupling and Magnetic Transition in MnBi$_2$Te$_4$/MnBr$_3$ Heterostructure

Yuanhao Zhu[1], Xixi Yuan[1], Ying Zhao[1], Jin Zhang[2], Zijing Ding[3, *], Huixia Fu[1, 4, †]

[1] *Center of Quantum Materials and Devices, Chongqing University, Chongqing 401331, China*
[2] *Laboratory of Theoretical and Computational Nanoscience, National Center for Nanoscience and Technology, Chinese Academy of Sciences. Beijing 100190, China*
[3] *Beijing Computational Science Research Center, Beijing 100193, China*
[4] *Chongqing Key Laboratory for Strongly Coupled Physics, Chongqing University, Chongqing 401331, China*

Bulk MnBi$_2$Te$_4$ (MBT) is an intrinsic antiferromagnetic topological insulator. However, its low Néel temperature of ~25 K severely restricts its practical applications. Here, we propose a van der Waals heterostructure composed of monolayer MBT (ML-MBT) and monolayer MnBr$_3$, an intrinsic Chern insulator possessing a high Curie temperature ($T_C$~200 K). By employing density functional theory calculations and Monte Carlo simulations, we demonstrate that interfacing ML-MBT with MnBr$_3$ significantly enhances the $T_C$ of ML-MBT by a factor of four to five. Electronic correlations characterized by the Hubbard parameter $U_2$ for Mn-$d$ orbitals in MnBr$_3$ play a crucial role in governing magnetic coupling within the system. At a moderate correlation strength of $U_2$ = 3.0 eV, slight structural distortions in MnBr$_3$ break intralayer symmetry, enabling robust interlayer ferromagnetic coupling and yielding a single, unified magnetic transition. Increasing $U_2$ reduces these structural distortions, weakens interlayer coupling, and induces two distinct magnetic transitions, indicating interlayer magnetic decoupling. Thus, the MBT/MnBr$_3$ heterostructure offers a novel approach for controlling magnetic order and enhancing the performance of spintronic devices.

## I. INTRODUCTION

Two-dimensional (2D) materials provide an ideal platform for exploring intriguing physical phenomena and serve as a rich source of candidates for the development of next-generation functional devices [1-4]. The discovery of long-range magnetic order at finite temperatures in monolayer CrI$_3$ [5,6] and bilayer CrGeTe$_3$ [7] has particularly stimulated extensive research into 2D van der Waals (vdW) magnets [8-11]. Prominent examples include ferromagnetic Fe$_3$GeTe$_2$ [12], antiferromagnetic FePS$_3$ [13,14], and room-temperature ferromagnetic VSe$_2$ [15]. Despite these advancements, the practical application of 2D magnets faces significant challenges. According to the Mermin-Wagner theorem [16], substantial magnetic anisotropy energy (MAE) is essential to suppress thermal fluctuations and stabilize magnetic order in two dimensions. Additionally, elevating the $T_C$ is crucial for the practical implementation of devices [17], prompting intensive efforts to enhance magnetization, discover new materials, and tune magnetic properties via external fields [18-20].

Recently, the antiferromagnetic vdW layered material MnBi$_2$Te$_4$ (MBT) has garnered significant attention due to its intrinsic magnetic topological insulating properties [21,22]. MBT comprises seven-layer (SL) units, with adjacent SL layers coupled antiferromagnetically. MBT thin films exhibit diverse topological phenomena, including the quantum anomalous Hall effect, Weyl semimetal states, and axion insulator states, making MBT a promising candidate for spintronic applications. However, monolayer MBT (ML-MBT) exhibits trivial topological properties, with a substantial indirect bandgap of 321 meV [23], which hinders topological phase transitions. Furthermore, ML-MBT displays nearly isotropic magnetic anisotropy [24] and possesses a very low $T_C$ of ~12 K, significantly limiting its practical applicability. Enhancing the $T_C$ and modulating quantum properties of MBT remains a critical challenge. Many attempts have been made to improve the performance of ML-MBT. For example, applying an external electric field can induce a topological phase transition in ML-MBT, raising its $T_C$ to 61 K [25]. Typically, combining two 2D materials into a heterostructure can modulate the overall performance [26-29], providing a platform for studying interfaces and device applications [30,31]. An MBT/In$_2$Se$_3$ heterostructure constructed using a 2D ferroelectric substrate can increase the $T_C$ of ML-MBT to 32 $K$ under compressive strain [32]. When the ML-MBT slab is combined with the 2D ferromagnetic semiconductor CrI$_3$, a quantum anomalous Hall state with high Chern number can be realized [33]. Recent studies have demonstrated that MnBi$_2$Te$_4$/XBi$_2$Te$_4$/MnBi$_2$Te$_4$ sandwich heterostructures (where X = Ge, Sn, Pb) can achieve a magnetic transition temperature of 38 K while preserving the quantum anomalous Hall effect [34].

In this study, we propose an MBT/MnBr$_3$ heterostructure that combines ML-MBT with monolayer MnBr$_3$, a ferromagnetic topological material characterized by a high $T_C$ of ~200 K [35]. Through first-principles calculations and Monte Carlo (MC) simulations, we investigate the electronic and magnetic properties of this heterostructure. Our results reveal a robust interlayer ferromagnetic coupling and a substantial enhancement of $T_C$ for ML-MBT, resulting in a four- to fivefold increase. We identify electronic correlations, characterized by the Hubbard parameter $U_2$ of Mn-$d$ orbitals in MnBr$_3$, as crucial factors governing magnetic coupling. At a moderate $U_2$ value (3.0 eV), slight structural distortions in MnBr$_3$ induce symmetry breaking, promoting strong interlayer coupling and resulting in a unified $T_C$ = 72 K. Conversely, increasing $U_2$ suppresses structural distortions, weakens interlayer magnetic coupling, and leads to magnetic phase separation, indicated by two distinct magnetic transitions ($T_{C1}$ = 56 K, $T_{C2}$ = 158 K). Additionally, through the analysis of the projected density of states (PDOS), charge transfer, potential energy diagrams, and atomic structure distortions, we elucidate the distinct magnetic exchange mechanisms dominated by two triangular regions, which exhibit a seesaw behavior in magnetic interactions. Our findings demonstrate a route to manipulating magnetic properties through electronic correlations, providing valuable theoretical insights for the development of advanced spintronic heterostructures. Our findings demonstrate a clear route to manipulating magnetic properties through electronic correlations, and also suggest an alternative application perspective in which magnetic transition behaviors could be utilized to probe microscopic electronic correlations.

## II. METHOD

Density functional theory (DFT) calculations of electronic structures and magnetic properties for the MBT/MnBr$_3$ heterostructure were performed using the Vienna Ab initio Simulation Package (VASP) [36] and the projected augmented wave (PAW) method [37]. The generalized gradient approximation (GGA) within the Perdew-Burke-Ernzerhof (PBE) [38] framework was employed for the exchange-correlation functional. A plane-wave basis set with a cutoff energy of 400 eV was adopted, and the Brillouin zone integration utilized a $\Gamma$-centered 3×3×1 $k$-point mesh. The atomic structure is fully relaxed, and the convergence criteria are that the total energy is less than $1\times10^{-7}$ eV and residual atomic forces are below $1\times10^{-3}$ eV/Å. A vacuum spacing exceeding 15 Å along the $z$-axis direction was introduced to prevent spurious interactions between adjacent periodic slabs. To better describe the Coulomb interaction of the localized 3$d$ orbitals of Mn atoms in the heterostructure, the GGA+U method [39] is employed. The structural relaxation was performed separately for each value of $U$. Interlayer vdW interactions between ML-MBT and MnBr$_3$ were described using the DFT-D3 correction method [40]. Spin-orbit coupling (SOC) was fully included in all self-consistent electronic structure calculations and PDOS analyses.

An effective spin Hamiltonian was constructed using the PASP [41] program via a machine learning-based approach. In the calculations, magnetic exchange interactions were considered for atomic pairs with separations less than 21 Bohr. To achieve accurate parameterization, a total of 500 random spin configurations were generated, and their energies were calculated using the DFT method. From these, 400 configurations were used as a training dataset, and the remaining 100 configurations were reserved for testing the effectiveness of the Hamiltonian model. MC simulations were performed with a Metropolis algorithm implemented in PASP to determine the $T_C$. Simulations employed a 9×9×1 supercell of the heterostructure, with each temperature point consisting of 1,000 MC equilibration steps followed by 500 thermalization steps. The subsequent 100 MC block steps were utilized for statistical averaging and analysis.

## III. RESULTS
### A. Atomic structure and interfacial magnetic coupling in MBT/ MnBr$_3$ heterostructure

ML-MBT is a trivial insulator with a ferromagnetic (FM) ground state and a space group of $P\bar{3}m1$ (No. 164). Monolayer MnBr$_3$ is an intrinsic Chern insulator [42] with a FM ground state and a narrow band gap, characterized by a space group of $P\bar{3}1m$ (No. 162). The optimized in-plane lattice constants of ML-MBT, with and without vdW correction, are 4.30 Å and 4.36 Å, respectively, which are in good agreement with previous reported results [43]. Given the critical role of vdW interactions in the heterostructure, we employ the lattice constant of 4.30 Å for our calculations. The MBT/MnBr$_3$ heterostructure is constructed by combining a 3×3 MBT supercell ($a$ = $b$ = 12.90 Å) and a 2×2 MnBr$_3$ supercell ($a$ = $b$ = 13.14 Å), compressing MnBr$_3$ by approximately 1.89% to match the ML-MBT lattice. The heterostructure exhibits $C_3$ symmetry and a space group of $P3$ (No. 143).

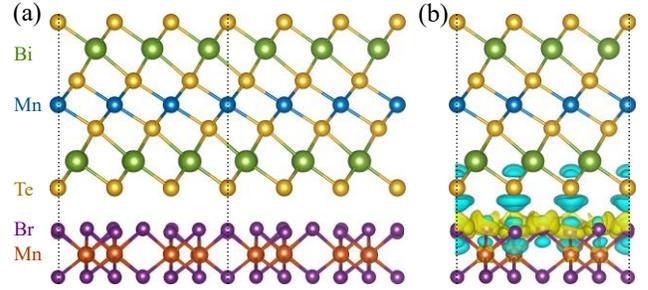

FIG 1. (a) Side view of the MBT/ MnBr$_3$ heterostructure, illustrating the magnetic Mn atoms in MBT (blue) and MnBr$_3$ (orange). The unit cell of the heterostructure is indicated by gray dashed lines, with lattice constants $a$ = $b$ = 12.90 Å. (b) Charge density difference plot showing the transfer of electrons from the MBT layer to the MnBr$_3$ layer. The cyan region indicates a decrease in charge density, while the yellow region represents an increase in charge density. The isosurface level is set to 0.002 $e$/Å³.

The atomic arrangement of the heterostructure is shown in Fig. 1(a). Although both ML-MBT and MnBr$_3$ adopt a hexagonal lattice, their magnetic structures differ. The Mn atoms in ML-MBT form a hexagonal close-packed (HCP) lattice, whereas those in MnBr$_3$ adopt a honeycomb configuration. Analysis of the charge density difference, as shown in Fig. 1(b), reveals electron transfer from ML-MBT to MnBr$_3$, resulting in an internal electric field at the interface. This electron redistribution primarily involves Bi and Te atoms near the ML-MBT interface and Mn and Br atoms adjacent to the MnBr$_3$ interface, facilitating interlayer charge neutrality. The distinct valence electron configurations—Mn$^{2+}$ (3$d^5$) in ML-MBT and Mn$^{3+}$ (3$d^4$) in MnBr$_3$—critically influence the interlayer magnetic coupling within the heterostructure. According to the electron-counting rule proposed in Ref. [44], monolayers with fewer than five electrons in their *d*-orbitals (type-I, $d^n$ with n < 5) favor interlayer ferromagnetic (FM) coupling when interfaced with type-II layers ($d^n$ with n ≥ 5). Thus, the ML-MBT (type-II) and MnBr$_3$ (type-I) interfaces naturally promote interlayer FM order. Moreover, their integration into a heterostructure may strengthen the interlayer magnetic coupling, leading to enhanced and stable FM interactions.

We separately performed first-principles calculations for the individual ML-MBT and MnBr$_3$ monolayers composing the heterostructure. In this work, the Hubbard *U* parameters for Mn in ML-MBT and MnBr$_3$ are denoted as $U_1$ and $U_2$, respectively. As shown in Supplementary Figs. S1 and S2, the atomic structure and electronic properties of ML-MBT remain robust against variations in $U_1$. In contrast, monolayer MnBr$_3$ exhibits significant sensitivity to the Hubbard parameter $U_2$. This sensitivity arises from the hybridization between Mn-3*d* and Br-*p* orbitals, which introduces substantial contributions from Mn-3*d* orbitals at the Fermi level, as illustrated in Supplementary Figs. S3 and S4. To achieve lattice matching within the heterostructure, we applied a moderate biaxial compressive strain of approximately 1.89% to monolayer MnBr$_3$. The calculations under this strain condition, shown in Supplementary Figs. S5 and S6, yield results consistent with those obtained using the pristine unit cell. Therefore, careful consideration and precise tuning of $U_2$ are essential for accurately capturing the electronic structure and magnetic behavior of the MnBr$_3$ layer.

Given that experimental realizations of heterostructures, such as the EuS/TI [45] and CrSe/TI [46] systems are often affected by interfacial disorder or chemical intermixing, we conducted a stability assessment of our proposed heterostructure, as summarized in Tables S1-S3. In these calculations, the Hubbard parameters were fixed at $U_1$ = 5.0 eV and $U_2$ = 5.0 eV. The binding energy indicates that the heterostructure is relatively stable, as energy increases are observed under both interlayer sliding and variations in interlayer spacing, suggesting that the system resides in a lowest-energy stable state. In our calculations, the chosen structural configuration was primarily aimed at preserving the $C_3$ symmetry of the system. This choice offers two clear advantages: (1) the high-symmetry structure significantly reduces computational complexity; and (2) by minimizing the number of exchange parameters, it simplifies the Heisenberg model, thereby facilitating the analysis of magnetic interactions.

## B. Magnetic interaction and $T_C$ in MBT/MnBr$_3$ heterostructure

To systematically investigate the magnetic properties of the heterostructure, we constructed an effective spin Hamiltonian based on spin invariants selected via first-principles calculations combined with machine-learning fitting. Subsequently, the magnetic transition temperature was determined by performing MC simulations with simulated annealing. To validate our approach, we initially calculated the magnetic exchange coefficients for the ML-MBT system, obtaining nearest-neighbor $J_1$ = -1.387 meV and next-nearest-neighbor $J_2$ = 0.089 meV, yielding a $T_C$ of ~13 K, which is in agreement with the results reported in Ref. [47].

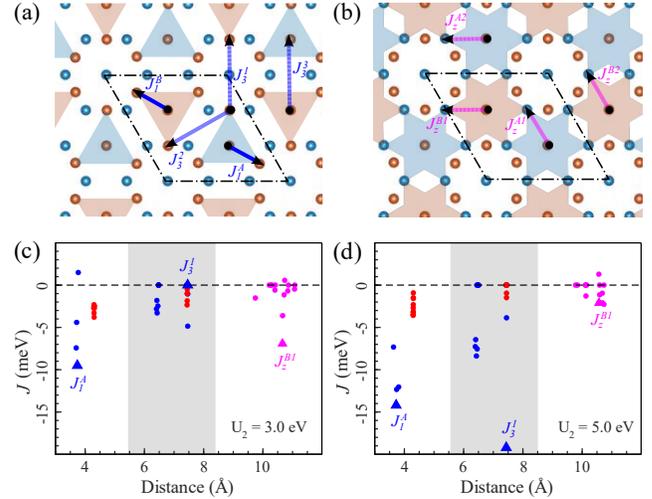

FIG 2. (a) Schematic diagrams of intralayer magnetic exchange interactions $J^{intra}$, where the light blue and light brown triangular regions represent regions *A* and *B*, respectively. The intralayer nearest-neighbor exchange $J_1$ corresponds to the exchange interaction between the Mn at the center of the triangular region and its nearest-neighbor Mn. The third-nearest-neighbor intralayer exchanges are further divided into two types: $J_3^{1(2)}$ and $J_3^3$. (b) Schematic diagrams of interlayer magnetic exchange interactions ($J^{inter}$), where two sets of triangular regions stack to form a hexagonal region and the interlayer exchange coefficients are labeled as $J_z^{A(B)1}$ and $J_z^{A(B)2}$, where the superscripts *A* and *B* correspond to the previously defined light blue and light brown regions, respectively. For simplicity, only the magnetic Mn atoms are shown in the figure, with all non-magnetic atoms and atomic bonds omitted. (c, d) Statistical distribution of magnetic exchange interactions (*J*) as a function of the interatomic distances in the MBT/MnBr$_3$ heterostructure, calculated using Hubbard parameters (c) $U_2$ = 3.0 eV and (d) $U_2$ = 5.0 eV, while $U_1$ = 5.0 eV is fixed. The red, blue, and pink dots represent $J^{MBT}$, $J^{MnBr_3}$ and $J^{inter}$, respectively. Points that significantly influence the magnetic properties of the heterostructure are highlighted with triangles.

For the heterostructure system, the following form of the Heisenberg Hamiltonian is considered (1):

$$H = H_0 + \sum_{ij} J_{ij}^{MBT} S_i \cdot S_j + \sum_{ij} J_{ij}^{MnBr_3} S_i \cdot S_j + \sum_{ij} J_{ij}^{inter} S_i \cdot S_j$$

$$= H_0 + H_{MBT} + H_{MnBr_3} + H_{inter} \quad (1)$$

where $H_0$ is the nonmagnetic part. The terms $J^{MBT}$, $J^{MnBr_3}$ and $J^{inter}$ represent the Heisenberg exchange interactions within the ML-MBT layer, within the MnBr$_3$ layer, and across the interface between these two layers, respectively. A negative *J* value indicates FM coupling, while a positive *J* value indicates AFM coupling. To clearly define the superscripts and subscripts associated with each magnetic exchange interaction in this work, each Hamiltonian can be expressed in the form of an effective Hamiltonian as follows (see Supplemental Material, Sec. I: Introduction):

$$H_{MBT} = \sum_{\mu} J_a^{\mu} h_{\mu} + \sum_{\nu} J_b^{\nu} h_{\nu},$$
$$(\mu = 1,2,\ldots,9; \nu = 1,2,\ldots,9) \quad (2)$$

Here, $J_a^{\mu}$ and $J_b^{\nu}$ represent the nearest-neighbor and next-nearest-neighbor magnetic exchange interaction terms within the ML-MBT layer, respectively. The $h$ denotes the basis functions used in the linear combination.

$$H_{MnBr_3} = \sum_{\mu} J_1^{\mu} h_{\mu} + \sum_{\nu} J_2^{\nu} h_{\nu} + \sum_{\lambda} J_3^{\lambda} h_{\lambda},$$
$$(\mu = 1,2,\ldots,4; \nu = 1,2,\ldots,8; \lambda = 1,2,\ldots,4) \quad (3)$$

Here, $J_1^{\mu}$, $J_2^{\nu}$ and $J_3^{\lambda}$ represent the nearest-neighbor, next-nearest-neighbor, and third-nearest-neighbor interaction terms within the MnBr$_3$ layer, respectively.

$$H_{inter} = \sum_{\kappa} J_z^{\kappa} h_{\kappa}, (\kappa = 1,2,\ldots,14) \quad (4)$$

Here, $J_z^{\kappa}$ represents the magnetic exchange interaction term between heterostructure layers. Subsequently, we obtained the magnetic exchange interactions of the heterostructure through DFT calculations and machine learning screening. The results are shown in Figures 2(c) and 2(d), where $U_1 = 5.0$ eV is fixed. The red, blue, and pink colors in the figures represent $J^{MBT}$, $J^{MnBr_3}$ and $J^{inter}$, respectively. Clearly, the negative interlayer exchange interactions confirm that the interlayer coupling of the heterostructure is FM, consistent with the earlier conclusions drawn from electron-counting analysis. It can also be observed that the magnetic exchange interaction $J_{MBT}$ within the ML-MBT layer is relatively small, thus having a minimal impact on the heterostructure.

Table I. The values of magnetic exchange interactions in the MBT/MnBr$_3$ heterostructure with different $U$, as chosen from Fig. 2, are listed. Negative values represent FM interactions, while positive values indicate AFM interactions. The subscripts 1, 3 and z refer to the intralayer nearest-neighbor, intralayer third-nearest-neighbor and interlayer exchange, respectively. The superscripts $A$ and $B$ indicate regions $A$ and $B$, respectively. The values represented as "--" correspond to negligible interactions ($< 1 \times 10^{-2}$ meV).

| Magnetic exchange coefficient | $U_2 = 3.0$ eV (meV) | $U_2 = 5.0$ eV (meV) |
| --- | --- | --- |
| $J_1^A$ | -9.51 | -14.21 |
| $J_1^B$ | -4.41 | -12.33 |
| $J_3^1$ | -- | -19.21 |
| $J_3^2$ | -4.86 | -- |
| $J_3^3$ | -1.04 | -3.86 |
| $J_z^{A1}$ | -- | -- |
| $J_z^{A2}$ | -0.47 | -2.28 |
| $J_z^{B1}$ | -6.91 | -2.09 |
| $J_z^{B2}$ | -3.61 | -0.96 |

We have listed the values for all the points shown in Figures 2(c) and 2(d) in Tables S9-S14 of the Supplementary Materials. This paper primarily focuses on the most significant magnetic exchange term, $J$, which is highlighted using triangles in Figures 2(c) and 2(d). The top view of the heterostructure is shown in Fig. 2(a), where only the magnetic Mn atoms are depicted. The MnBr$_3$ supercell contains eight Mn atoms, two of which are vertically aligned with the ML-MBT layer. These two Mn atoms serve as the centers of two triangular regions, labeled as regions $A$ and $B$, corresponding to the intralayer exchange terms $J^A$ and $J^B$, respectively. Fig 2(b) illustrates the interlayer exchange terms for these two regions. Table I summarizes the values of the key magnetic exchange interactions calculated under different $U_2$ configurations. Based on Fig. 2 and Table I, increasing $U_2$ from 3.0 eV to 5.0 eV strengthens the intralayer interaction $J_1^A$ from -9.51 meV to -14.21 meV dominated by region $A$, while simultaneously weakening the interlayer interaction $J_z^{B1}$ from -6.91 meV to -2.09 meV dominated by region $B$. This inverse correlation between intralayer and interlayer magnetic interactions exhibits a seesaw-like behavior.

We have also included the results for other $U$ configurations in Figure S8. Upon comparison, we found that the magnetic interaction within the MBT/MnBr$_3$ heterostructure is more sensitive to changes in $U_2$, but more robust to changes in $U_1$. This is consistent with the previous analysis, and therefore, in the subsequent calculations, we fixed $U_1 = 5.0$ eV and only varied the parameter $U_2$.

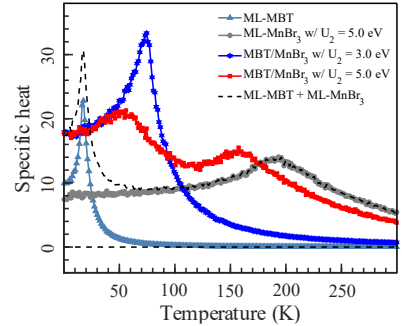

FIG 3. Temperature-dependent heat capacity curves obtained from parallel annealing Monte Carlo simulations. The Hubbard parameter $U_1 = 5.0$ eV is fixed for all simulations. Magnetic transition temperatures are identified by the inflection points in the curves. The red and blue curves correspond to the MBT/MnBr$_3$ heterostructure under different $U_2$ configurations. The indigo and gray curves represent the transitions of isolated monolayers of ML-MBT and MnBr$_3$, respectively. The black dashed line shows the numerical sum of the heat capacity contributions from the isolated ML-MBT and MnBr$_3$ without interlayer coupling.

The variation in magnetic exchange interactions reflects differences in the temperature-dependent heat capacity curves. The magnetic transition of the heterostructure, as determined via MC simulations, is presented in Fig. 3, highlighting several notable phenomena. The Curie temperature reaches $T_C = 72$ K with $U_2 = 3.0$ eV, a fivefold increase compared to ML-MBT. However, as $U_2$ increases to 5.0 eV, significant phase separation emerges, and the magnetic transition temperature splits into two distinct values: $T_{C1} = 56$ K and $T_{C2} = 158$ K. More detailed evidence of this phase separation is provided in Supplementary Fig. S7, which shows the temperature dependence of the average magnetic moments for the ML-MBT and MnBr$_3$ layers separately. It clearly demonstrates that at $U_2 = 5.0$ eV, the two layers lose their ferromagnetic ordering sequentially upon heating, transitioning to paramagnetic states at distinct temperatures. This behavior explicitly indicates a decoupling of their magnetic phase transitions.

The magnetic phase separation primarily originates from the variation in interlayer magnetic exchange interactions. In Fig. 3, the indigo and gray curves represent the magnetic transition temperatures of isolated ML-MBT and ML MnBr$_3$. The black dashed line represents the sum of these two curves, excluding the interlayer magnetic exchange terms, which contrasts with the heterostructure. As discussed previously, the interlayer magnetic exchange $J^{inter}$ for $U_2 = 3.0$ eV is stronger than that for $U_2 = 5.0$ eV. Consequently, as $J_{inter}$ decreases, the single $T_C$ splits into two distinct peaks, and the separation between these peaks progressively widens, evolving from the blue curve, through the red curve, to the black dashed line (Fig. 3). This result identifies that the strength of interlayer coupling tuned by $U_2$ determines whether the MBT/MnBr$_3$ heterostructure exhibits strong magnetic coupling or magnetic phase separation.

### C. Electronic correlation mediated structural distortion and interlayer magnetic coupling

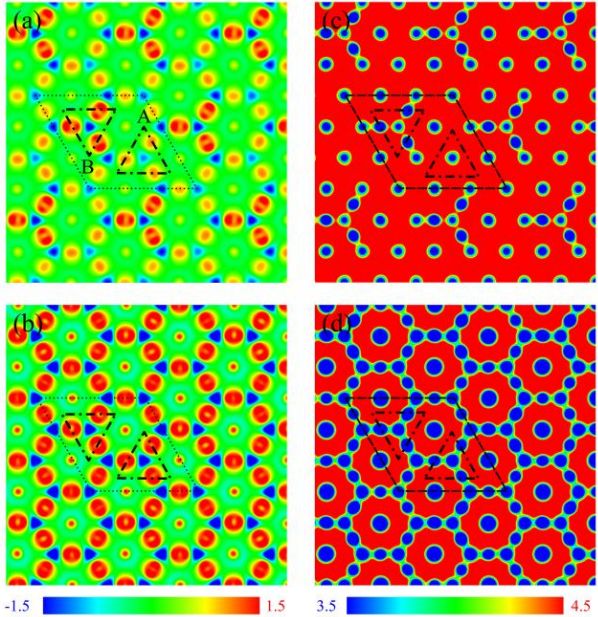

FIG 4. Charge transfer maps with (a) $U_2 = 3.0$ eV and (b) $U_2 = 5.0$ eV. The unit is $10^{-3}$ $e/\text{Å}^3$. Corresponding local potential distribution maps for (c) $U_2 = 3.0$ eV and (d) $U_2 = 5.0$ eV, respectively. The unit is eV. The black triangles outline two distinct triangular regions, labeled as regions $A$ and $B$. The maps depict cross-sections taken along the (001) plane at the maximum local potential position between the two layers of the heterostructure. The gray dashed line marks the calculated unit cell of the heterostructure.

The interlayer charge transfer in the heterostructure unambiguously reflects the strength of the interlayer magnetic coupling. To support this, we present the spin-resolved total density of states (TDOS) projected along the $S_z$ direction for the heterostructure in Fig. S16. The results clearly show that the charge transfer near the Fermi level is predominantly contributed by spin-up states. To further illustrate this, Fig. S17 shows the spatial distributions of partial charge density near the Fermi level under different energy windows, again confirming that the dominant contribution comes from the spin-up channel. To explore the magnetic implications of this charge transfer, we calculated the averaged partial charge distribution along the $z$-direction near the Fermi level, as shown in Fig. S18. As the magnetic exchange between Mn atoms is mediated by orbitals near the Fermi energy, we find that the partial charges associated with Mn atoms, highlighted by the light red and blue dashed lines in Fig. S18c, decrease with increasing $U_2$. This trend reflects a weakening of interlayer magnetic exchange due to charge redistribution.

This analysis only addresses the averaged effect of charge transfer at a macroscopic level. To further elucidate how electron correlations in the MnBr$_3$ layer, characterized by the parameter $U_2$, influence both interlayer and intralayer magnetic interactions, we performed a detailed investigation of structural distortions, local electrostatic potentials, and charge transfer at the interface.

Table II. The bond angles $\theta$ (Mn-Br-Mn) and bond lengths $d$ (Mn-Br) within the MnBr$_3$ layer of the MBT/MnBr$_3$ heterostructure. Subscripts $A$ and $B$ denote the two triangular regions. The superscripts 1 and 2 represent exchange paths through the lower and upper Br atoms, respectively. The superscripts L and R indicate bond lengths closer to and farther from the central Mn atom.

|  | $U_2 = 3.0$ eV | $U_2 = 5.0$ eV |
| --- | --- | --- |
| $\theta_A^1$ (°) | 92.79 | 90.51 |
| $\theta_A^2$ (°) | 91.33 | 87.95 |
| $\theta_B^1$ (°) | 92.38 | 91.09 |
| $\theta_B^2$ (°) | 88.00 | 88.10 |
| $d_A^{1L}/d_A^{1R}$ (Å) | 2.61/2.54 | 2.62/2.62 |
| $d_A^{2L}/d_A^{2R}$ (Å) | 2.68/2.54 | 2.68/2.68 |
| $d_B^{1L}/d_B^{1R}$ (Å) | 2.65/2.49 | 2.62/2.61 |
| $d_B^{2L}/d_B^{2R}$ (Å) | 2.60/2.74 | 2.68/2.70 |

We choose the equilibrium interfacial plane between the ML-MBT and MnBr$_3$ layers as the position of maximum local potential, indicated by the dashed line in the z-directional potential distribution maps, shown in Fig. S9(a) and S9(b). Cross-sectional maps of charge transfer and local potential at this equilibrium plane are depicted in Figs. 4. For $U_2 = 3.0$ eV, asymmetry between regions $A$ and $B$ emerges, clearly signaled by distinct charge transfer patterns. This asymmetry reflects slight structural distortions within the MnBr$_3$ layer, which directly affect the local electronic environments and, consequently, the magnetic interactions. Such distortion-induced asymmetry is also evident in the local potential distribution shown in Fig. 4(c).

In addition, Fig. S10 provides a detailed top-view analysis of the MnBr$_3$ layer, distinguishing between upper and lower Br atoms and quantifying variations in Mn-Br bond lengths and Mn-Br-Mn bond angles, listed in Table II. Compared to the symmetric structure maintained at $U_2 = 5.0$ eV, the symmetry between the central Mn atom and its neighboring Mn atoms with the two triangular regions is clearly broken at $U_2 = 3.0$ eV, confirming the presence of structural distortions. These subtle lattice distortions enhance interlayer magnetic coupling, leading to a unified ferromagnetic transition.

Moreover, we have summarized the magnetic moments of Mn atoms in monolayer MBT, monolayer MnBr$_3$, and the heterostructure in Tables S7 and S8. The magnetic moment of MBT is reduced after forming the heterostructure, while the MnBr$_3$ layer exhibits an increased moment. These results support our conclusion that electrons transfer from ML-MBT to MnBr$_3$, and this transfer is spin-polarized in the spin-up channel. Additionally, we observe enhanced magnetic moment fluctuations in MnBr$_3$ under $U_2 = 3.0$ eV. We attribute this to symmetry breaking and enhanced interlayer magnetic coupling, which is consistent with the analysis presented above.

Structural distortions affect the PDOS distribution, as shown in Fig. S11. Focusing particularly on the dominant interlayer exchange term, $J_z^{B1}$, we identify the key atomic orbitals along the interlayer super-super-exchange path. As shown in Fig. S11(a), variations in the local density of states (LDOS) between the central and nearest Mn atoms become pronounced at $U_2 = 3.0$ eV. Enlarged LDOS plots in Figs. 5(a)-(b) further highlight differences between regions $A$ and $B$, as well as between central and neighboring Mn atoms within each region. These LDOS disparities result in energy level splitting, closely correlating with the observed structural distortions and robust interlayer magnetic coupling.

To quantitatively describe this energy level splitting, we define $\Delta dos$ = LDOS (central Mn) − LDOS (nearest Mn), where LDOS (central Mn) and LDOS (nearest Mn) refer to the LDOS of the central Mn and nearest Mn atoms, respectively. Calculations were performed across a range of $U_2$ values, from 3.0 to 5.0 eV, with increments of 0.2 eV, revealing a clear evolution as shown in Figs. 5(c) and 5(d). As $U_2$ increases, the energy level splitting steadily diminishes, ultimately vanishing around 3.8 eV. This reduction in splitting coincides with weakened structural distortion and diminished interlayer magnetic coupling, culminating in a magnetic decoupling characterized by two separate Curie temperatures.

## IV. SUMMARY

In summary, we systematically investigated the structural, electronic, and magnetic properties of the MBT/MnBr$_3$ heterostructure using first-principles calculations, machine learning fitting, and MC simulations. Our results indicate that the $T_C$ of ML-MBT can be enhanced by a factor of four to five through heterostructure formation. The strength of interlayer magnetic coupling, modulated by electronic correlations in the MnBr$_3$ layer described by the Hubbard parameter $U_2$, depends sensitively on structural distortions. With $U_2 = 3.0$ eV, symmetry-breaking distortions promote stronger interlayer ferromagnetic coupling, resulting in a unified magnetic transition. Conversely, larger $U_2$ values reduce these distortions and weaken the interlayer coupling, leading to magnetic phase separation characterized by two distinct $T_C$ values (56 K and 158 K). Moreover, considering the inherent challenge of precisely determining the Hubbard parameter from first-principles calculations, our findings suggest that magnetic transitions could serve as practical experimental probes to investigate electronic correlations in complex magnetic systems.

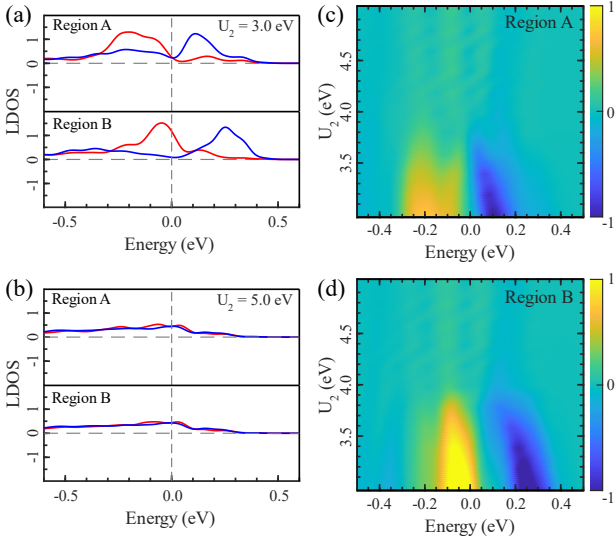

FIG 5. Enlarged LDOS plots for region $A$ (upper) and region $B$ (lower) for the $U_1 = 5.0$ eV configuration with (a) $U_2 = 3.0$ eV and (b) $U_2 = 5.0$ eV. The red and blue curves represent the LDOS of central Mn and nearest Mn atoms of each region, respectively. (c) and (d) The difference ($\Delta dos$) in LDOS for triangular regions $A$ and $B$ with $U_2 = 3.0, 3.2, \ldots, 4.8, 5.0$ eV. The $\Delta dos$ is defined as $\Delta dos$ = LDOS (central Mn) – LDOS (nearest Mn), quantifying the degree of energy level splitting. The linear interpolation method was applied, inserting nine points between adjacent $U_2$ configurations to create a continuous image. The color bar represents the difference between the LDOS (i.e., $\Delta dos$). The LDOS in the figures above is all projections of the density of states in the $z$-direction.

## ACKNOWLEDGMENTS

The authors acknowledge financial support from the National Natural Science Foundation of China (Grant Nos. 12104072, 12347101). Numerical computations were partly performed at the Hefei advanced computing center.

---

† hxfu@cqu.edu.cn
* zding@csrc.ac.cn